# PolicyCLOUD: A prototype of a Cloud Serverless Ecosystem for Policy Analytics

Ofer Biran[1], Oshrit Feder[1], Yosef Moatti[1], Athanasios Kiourtis[2], Dimosthenis Kyriazis[2], George Manias[2], Argyro Mavrogiorgou[2], Nikitas M. Sgouros[2], Martim Taborda Barata[3], Isabella Oldani[3], María Angeles Sanguino[4], Pavlos Kranas[5]

```
Email: {biran, oshritf, moatti}@il.ibm.com, {kiourtis, dimos, gmanias,
   margy, sgouros}@unipi.gr, {martim.tabordabarata, isabella.oldani}
        @ictlc.com, maria.sanguino@atos.net, pavlos@leanxcale.com
```

## Abstract

We present PolicyCLOUD, a prototype for an extensible, serverless cloud-based system that supports evidence-based elaboration and analysis of policies.

PolicyCLOUD allows flexible exploitation and management of policy-relevant dataflows by enabling the practitioner to register datasets and specify a sequence of transformations and/or information extraction through registered ingest functions. Once a possibly transformed dataset has been ingested, additional insights can be retrieved by further applying registered analytic functions.

PolicyCLOUD was built as an extensible framework toward the creation of an analytic ecosystem. As of now, we developed several essential ingest and analytic functions that are built-in within the framework. They include data cleaning, enhanced interoperability, and sentiment analysis generic functions.

PolicyCLOUD has also the ability to tap on the analytic capabilities of external tools. We demonstrate this with a Social Analytics tool implemented in conjunction with PolicyCLOUD and show how to benefit from policy modeling, design and simulation capabilities.

Furthermore, PolicyCLOUD has developed a first of its kind legal and ethical framework that covers the usage and dissemination of datasets and analytic functions throughout its policy-relevant dataflows.

The article describes and evaluates the application of PolicyCLOUD to four families of pilots that cover a wide range of policy scenarios.

## 1  Introduction

Policy practice is essentially eclectic, since policy makers can choose freely among a range of scientific methods, information sources and application methodologies to solve practical problems (Dunn, 2017). In fact, policy makers generate and/or rely on an ecosystem of different types of data sources and analytic methods. Each of these needs to be properly registered, filtered, analysed, validated, and searched, so that value can be extracted from those data and methods. To the best of our knowledge, there is no tool or platform today that efficiently and effectively supports the policy-making process in all of its unique aspects.

PolicyCLOUD, an ongoing EU Horizon 2020 project, seeks to deliver an innovative, cloud-based and data-centric approach to policy-making (Kyriazis, 2020). The project members are developing a unique integrated platform that targets the full lifecycle of policy-making: from creation through monitoring, analysis and compliance.

---

[1] IBM Research, Haifa, Israel

[2] University of Piraeus, Piraeus, Greece

[3] ICT Legal Consulting, Milan, Italy

[4] Atos Spain S.A., Research and Innovation Department, Madrid, Spain

[5] LeanXcale, Madrid, Spain

This article focuses on the PolicyCLOUD infrastructure and how we applied it to four families of pilot use cases: 1) RAD studies how policies can tackle the danger of radicalization 2) WINE is related to the (private) policies for marketing Aragon wine, 3) SOF deals with the analysis and elaboration of various policies for the Sofia municipality, ranging from air pollution to traffic related policies, 4) CAM relates to policy making for the London Camden district with regards to unemployment in this area.

This article describes the types of information sources and analytic capabilities supported by PolicyCLOUD, as well as their integration in a novel and extensible cloud-based framework. We refer to this cloud framework as the Data Acquisition and Analytics layer (DAA). The DAA controls the full data flow from data sources to the repository of the platform. It permits the registration and application of analytic functions during and after the ingestion of a dataset. We refer to the functions in the first case as "ingest functions" and to the rest as "analytic functions". Once a dataset is registered, PolicyCLOUD enables the application of functions which are stored in a library of functions that is maintained and constantly updated using DAA. New functions can be added to the library for use with a broad scope of datasets, or for use with a particular dataset. Thus, the DAA can be used to build a rich and flexible ecosystem of data and analytic methods. This ecosystem can be used to guide important aspects of policy-making, such as policy design, analysis, monitoring and compliance with ethical and legal requirements.

One advanced feature of PolicyCLOUD is its ability to exploit the power of simulations to model and reason about the outcomes of various alternatives during policy design. The framework includes a novel meta-simulation methodology that makes it easy to simulate and examine proposed policies, as well as compare the analysis and evaluation of their assumptions, mechanics and outcomes. In Section 6, we explain how this methodology provides insight and transparency to policy design and enhances the quality of debate and critique over policy creation.

One of the primary goals of this article is to share our experience and the knowledge acquired while developing this integrated framework and its individual components; it is also our aim to disseminate the best practises and lessons learned from our development efforts.

The structure of this article is as follows. Section 2 provides an overview of the open-source technologies on which the DAA is based. In Section 3, we detail the architecture of the DAA and the cloud gateway. Section 4 presents the legal and ethical concerns relevant to a platform like PolicyCLOUD. Section 5 details the generic ingest analytic technologies developed in PolicyCLOUD: data cleaning, enhanced interoperability and sentiment analysis. Section 6 details the social dynamics analytics which adds policy modelling and simulation capabilities to the framework. In Section 7, we share our evaluation results and lessons learned in terms of efficiency, adequacy and ease-of-use for integrated cloud-based data acquisition and analytics frameworks in policy-making contexts. Section 8 discusses the related work and we then conclude and point to what we believe should be the most important extensions of our work in Section 9.

## 2 Overview of Relevant Open-Source Technological Infrastructure

This section, briefly describes the cloud-based platform and open-source software tools we used to implement the DAA environment and why we found them most suitable. We chose to implement the DAA as a cloud-based serverless platform. This choice was natural since the cloud-native development model has two critical advantages: first, it is fully and automatically scalable, allowing developers to build and run applications without having to manage servers; secondly, this option allows for the platform to leverage a pay-as-you-use financial model which will typically be very advantageous considering the substantial fluctuation in the load of analytics activities which may be required by different platform users.

### 2.1 Kubernetes Cluster

The first design question raised by the DAA environment relates to the underlying virtualization platform: should it be based on virtual machines or on containers? We chose the container-based solution for a number of reasons. First, we saw advantages to its efficient application deployment and overall support for a continuous integration and delivery cycle.

Second, containers are the most appropriate for serverless platforms, and would be fully aligned with the extensibility and reusability requirements of the DAA. Third, containers are also the best fit for a microservices architecture, which enables easy deployment and separation of the DAA components. Fourth, using containers would offer the framework portability between cloud providers. Fifth, the growing popularity of containers and their strong ecosystem would be especially beneficial in open-source communities.

Once we decided to use a container-based solution, the Kubernetes (Kubernetes, n.d.) container management platform was a clear choice, being the leading open-source container management platform. Kubernetes is used in production as the base for private on-premises cloud systems for a growing number of enterprises, and is being offered by almost all cloud providers as a managed dedicated cluster. Kubernetes runs distributed applications resiliently, by handling scaling, load balancing and failover (e.g., automatically replacing containers that go down) Kubernetes also provides deployment patterns that drastically simplify application deployment and management.

## 2.2 OpenWhisk Cluster

Apache OpenWhisk (Apache OpenWhisk, n.d.) is an open-source, lightweight, serverless platform capable of deploying functions written in any language. OpenWhisk offers a simple programming model that allows function developers to concentrate on the function's logic, because the deployment and activation details are taken care of transparently by the platform. OpenWhisk uses containers to wrap functions, and can be deployed and integrated perfectly in a Kubernetes environment. Another important capability is that OpenWhisk allows functions to be activated by specified trigger events and execution rules, which would be perfect for a sequence of ingest functions.

## 2.3 Apache Kafka

Data streaming is the practice of (1) capturing data in real-time as events streams from sources, such as databases, cloud services, or software applications; (2) storing these event streams durably for later retrieval; (3) manipulating, processing, and reacting to event streams in real-time, as well as retrospectively; and (4) routing event streams to different destinations. The Apache Kafka (Apache Kafka, n.d.) data streaming platform is used for reliable data connectivity between components. The following are its key capabilities:

1. Publication (write) and subscription (read) to event streams, including the continuous importing/exporting of data from/to other systems;
2. Durable and reliable storage of event streams
3. Processing of event streams in real-time or retrospective.

kSQL (KSQLDB, n.d.) is an interesting extension that provides a streaming SQL engine running on top of Kafka. This allows us to continuously apply SQL queries to data channels and to route their output as sub-channels.

## 3 Architecture

### 3.1 Data Acquisition and Analytics Layer

The DAA is the central layer of PolicyCLOUD as it exploits the cloud infrastructure layer and provides the analytic API to the Policy layer which directly interfaces with the PolicyCloud users. It offers simple and efficient ways to a) register datasets and functions, b) apply ingest functions to pre-process and/or analyse data while it is ingested (streaming and non-streaming), such as transforming data or performing sentiment analysis on tweets, and c) apply analytic functions to stored data. Both the datasets and functions can be reused.

The DAA can also be used to manage legal and ethical concerns by requiring the data/analytic provider to enter relevant information explaining how such concerns have been dealt with. For example, this may include legal or contractual limitations on the use of datasets, algorithmic bias, and trade-offs. Any information submitted can be subsequently retrieved when the relevant artefact is listed; this enables potential users to decide on the legal or ethical adequacy of the dataset/function in an informed manner.

The DAA will typically be used by:

1. *Data providers* that manage the lifecycle of PolicyCloud's datasets through registration, deletion, and update. Upon registration, the raw data may be modified by a sequence of selected relevant ingest functions. Two primary categories of data sources are supported: a) streaming sources - datasets continuously ingested by the platform and b) sources at rest - static datasets that are ingested at once.
2. *Analytic providers* that similarly manage the lifecycle of general-purpose or specialised functions:
    a. Ingest functions to transform datasets. Such as a) removing unnecessary fields, b) extracting knowledge (e.g., sentiment from text) c) complying with legal/ethical requirements (e.g., by removing unnecessary personal information);
    b. Analytic functions, to be applied to ingested data.
3. *Policy practitioners/policy makers* who use PolicyCLOUD to apply analytic functions on datasets as appropriate to support their policy-making goals and decisions.

The DAA API gateway exposes the DAA functionality using a web interface, which is implemented as a set of serverless functions running in an OpenWhisk cluster. Data flows from the gateway over Apache Kafka until it ultimately gets stored in the platform data repository. Each function is executed in its own isolated environment (container), which is key for scalability and parallelism. The DAA includes a common Gitlab (GitLab, n.d.) structure storing the code registry of functions, as well as a common container registry for functions' docker (Docker, n.d.) images which package all the required dependencies. During function registration, files and images are automatically pulled from this Gitlab structure to create serverless functions.

The current list of reusable functions includes tools for: data cleaning, enhanced interoperability and sentiment analysis. In Section 5, we explain these components in detail, and show how policy makers can benefit from the DAA ecosystem and the implemented analytics technologies.

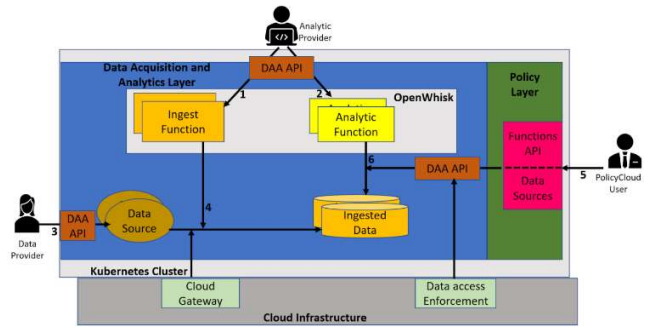

*Fig. 1 Architecture for Data Acquisition and Analytics Layer*

Fig. 1 depicts the DAA architecture and APIs. The analytic provider uses the DAA API (arrows 1 + 2) to register an ingest function or an analytic function. Administration privileges are required for both. Each function reads incoming data as a JSON string from a request message parameter and returns the transformed message as a JSON string. This output string may itself serve as input for a subsequent function, otherwise it will be stored in the PolicyCloud data store. A data provider uses the DAA API (arrow 3) to register a dataset by providing dataset information (metadata), the final schema, and optionally, the sequence of transformations to be applied to the dataset. Once the registration is invoked, the ingestion process of the dataset is automatically triggered and the dataset is stored in the DAA backend. This may occur after a sequence of ingest functions (e.g., data filtering/cleaning) are applied (arrow 4). At a later phase, a PolicyCLOUD user can apply a registered analytic function on the registered dataset (arrows 5 + 6).

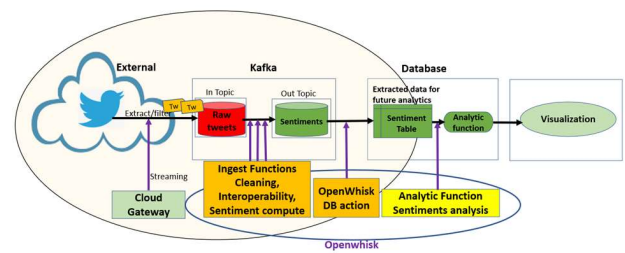

*Fig. 2 An example of streaming data path with sentiment ingest and analysis*

Fig. 2 provides an example of the workflow in which initial analytics are applied on streaming data. In this example, a social network (Twitter) has been registered as an information source, and a cloud gateway connector with specified filtering parameters (e.g., keywords) streams the data over Kafka, which has been integrated with OpenWhisk. The data cleaning, enhanced interoperability, and sentiment analysis ingest functions that have been registered for this dataset are invoked whenever new data is streamed. A sequence of additional ingest functions may be applied before the transformed data is written to the DAA data repository. Once stored, deeper analytics can be applied on the dataset as depicted by the "Analytic Function Sentiments Analysis".

## 3.2 Cloud Gateway

The DAA supports communication with the platform's cloud-based infrastructure through a cloud gateway and API component. The cloud gateway offers unified gateway capabilities that allow the transfer of streaming and batch data to the DAA. As the only data entry point for the DAA, the cloud gateway allows microservices to act cohesively and provide a uniform experience for each platform user.

The main goal of this component is to raise the invocation level by providing asynchronous request processing for these multiple microservices. This enables the acquisition of multimodal data from various information sources and aggregation of the results.

In addition, the cloud gateway supports client-side load balancing, which allows the platform to leverage complex balancing strategies such as caching, batching, and service discovery, while handling multiple protocols and implementing adapters for different technologies. On top of this, several mechanisms and microservices are used to check and evaluate that the provided raw data is in accordance with the data schema defined by the data provider.

Following the Gateway Pattern (Richardson, n.d.), a high-level open API specification is offered to the users, where each high-level action (e.g., function registration) typically invokes many low level microservices. The authentication mechanisms are applied at the gateway level. Since we are using the OAuth2 (OAuth 2.0, n.d.) protocol, the authentication server can be a separate component or even a third-party service.

# 4 Legal and Ethical Framework

## 4.1 Analytics Functions and Data Source Registration

Both ingest and analytic functions should be held to a high standard of legal/ethical compliance for a variety of reasons. One of these reasons is to ensure that the platform stays legal to use in the EU, for example, by preventing the registration/use of tools that do not meet applicable legal requirements. Another reason is to ensure the platform's data security, for example, by preventing the registration/use of tools that may compromise the platform's integrity, or the confidentiality of data stored on the platform. We also need to preserve the platform's trustworthiness, for example, by preventing the registration/use of tools that do not meet baseline ethical standards and thereby introduce a risk of deriving skewed, biased, inaccurate, or otherwise misleading information from the datasets to which they are applied. Lastly, this compliance can prevent the infliction of potential reputational damage on PolicyCLOUD users.

The PolicyCLOUD platform has also been designed to allow additional datasets to be registered by data providers. This registration allows us to implement measures that ensure the datasets can be lawfully leveraged by users, in the sense that their intended use is not subject to legal restrictions (e.g., contractual terms, database copyright protection, database *sui generis* right protection, copyright protection, privacy/data protection requirements). It may happen that a given dataset was built with inherent prejudice or biases, for example, where the data within does not properly represent the target population for which a given policy is designed, or by overrepresenting one gender or one ethnicity over another. In this case, policy decisions based on the datasets may ultimately reflect that same prejudice or bias, with potential unforeseen and unjustified harmful impact on individuals and communities. It is therefore important to ensure that the PolicyCLOUD platform can be lawfully leveraged by its users, at least within the EU legal framework, and to maximise user adherence to the platform and societal acceptability of policies created using the platform—and implement measures that address these risks.

A balance must be struck between maximising legal/ethical compliance on the one hand and avoiding overly restrictive registration processes for functions/datasets on the other. The former may ultimately compromise the effectiveness of the PolicyCLOUD platform, and the latter can trigger risks related to failure in meeting a high standard of legal/ethical compliance mentioned above.

We can help create this balance in advance by requiring that analytic/data providers document measures taken to address applicable legal/ethical requirements; this can be done through appropriate fields added to the platform's registration APIs. These include fields with details on specific measures taken to address the risk of biases inherent to a function/dataset, or other relevant legal/ethical constraints that may exist (e.g., the existence of personal data in a dataset, the management of relevant trade-offs in function development, authorisation from relevant rights holders). Guidelines on relevant measures to be documented are made available to analytic/data providers during the registration process. This helps those providers understand what information/documentation would be considered useful and valid. All analytics/data providers' input during registration will be linked, on the PolicyCLOUD platform, to the function/dataset and can later be accessed by any user. This should allow users to make informed and risk-based decisions on whether or not to leverage a given function/dataset, as well as to critically examine the output generated by the function/dataset in the context of their policy-making decisions.

## 4.2 Access Control

To ensure that access to specific functions and datasets is appropriately controlled, we developed a data governance and privacy enforcement mechanism, based on the Attribute-Based Access Control (ABAC) scheme. This includes a model and model editor used to define access policies and enforce them, and an ABAC authorisation engine, which is used to evaluate policies and attributes, thus enforcing protection and privacy-preserving policies.

To complement these technical controls, we put in place contractual limitations on any individual's ability to process personal data collected and managed via the PolicyCLOUD platform. For example, these include enforceable contractual obligations imposed on personnel with access to the platform's backend, and on platform users via the platform's Terms and Conditions.

## 4.3 Data Subject Rights Management

Because personal data may be included in datasets processed via the DAA, we need to ensure that the DAA allows for the exercise of data subjects' rights (e.g., under the EU General Data Protection Regulation/GDPR). The platform should, at the very least, not create any relevant technical obstacles to the exercise of these rights. As such, the PolicyCLOUD platform's design and implementation has been configured to make it capable of facilitating the exercise of these rights by having effective mechanisms in place to allow for this. This also considers the general right of individuals to an overview of, and easy access to, any of their personal data that may be processed via the platform; this may be relevant where ingested data sources are not fully aggregated/anonymised. To allow for this, the platform and the PolicyCLOUD data repository are being designed to allow the querying of any such data to identify, extract, modify, and/or delete personal data pertaining to an individual.

## 4.4 Data Configuration Management

Where personal data is included in a dataset to be processed via the PolicyCLOUD platform, any personal data collected and stored must be adequate, relevant, and limited to what is necessary in relation to the specific purpose(s) for which the dataset is to be processed. This requires the amount of personal data collected from each dataset to be minimised.

Moreover, any personal data stored on the platform should neither be stored in a form that permits the identification of the individuals to which it relates, nor should it be stored for any longer than necessary. In other words, specific retention periods should be defined for such data, considering the purposes for which such data are used; these retention periods are implemented in the PolicyCLOUD project and on the platform. When a retention period is exceeded, the personal data to which it relates must be erased or anonymised (e.g., via aggregation), unless its further retention can be legally justified. Furthermore, appropriate steps must be taken to verify the accuracy of any personal data collected and stored on the platform, and to maintain the accuracy of such personal data over time.

To address this, mandatory and optional data constraints can be defined on the PolicyCLOUD

platform to configure the parameters under which data validation, cleaning, and verification activities are carried out by ingest functions applied to the dataset. This gives data providers and PolicyCLOUD users control over the specific data points of a dataset that can be registered and leveraged via the platform. In particular, a dataset can be configured, as part of the platform's data cleaning processes (further explored in the next section), so that personal data is not collected or processed unnecessarily by the platform (e.g., configuring the platform so that, when ingesting a dataset, identifiers such as names, usernames, national ID numbers, IP addresses, dates of birth - are not collected or further processed). This enables unnecessary personal data to be removed from datasets prior to further processing via the platform, which provides greater assurances of privacy and data protection compliance, as well as of data quality (i.e., that only relevant and necessary data will be further processed on the platform).

## 5  PolicyCLOUD Ingest Analytics

This section details three ingest analytics technologies developed in PolicyCLOUD: data cleaning, enhanced interoperability, and sentiment analysis.

Data cleaning and enhanced interoperability are highly linked since during the data cleaning process inaccurate data is detected and corrected (or removed). It is then provided to the enhanced interoperability process for the extraction of semantic knowledge and the interlinking/correlation of the ingested data.

### 5.1  Data Cleaning

The goal of the data cleaning component is to ensure that all the data collected from possibly heterogeneous information sources will be as clean and complete as possible. Over the past decade, devices, organisations, and humans have begun to continuously produce and deal with data. Faster innovation cycles, improved business efficiencies, more effective research and development, and now policy making, are just a few of the benefits of efficiently using and understanding data (Gutierrez, 2020). All these create numerous challenges, including the challenge of volume, as well as the problem of generating insights in a timely fashion. Data cleaning can help facilitate the analysis of large datasets by reducing complexity and improving data quality.

Many authors have proposed data cleaning algorithms to remove noise and data inconsistencies, such as the work by Somasundaram et al. (Somasundaram, 2011). Another approach by Saqib et al. increases the efficiency of data warehouses is the creation of materialised views that pre-process and avoid complex resource intensive calculations (Saqib, 2012). Dallachiesa et al. (Michele Dallachiesa, 2013) proposed the NADEEF architecture which allows users to specify multiple types of data quality rules, that uniformly define what is wrong with the data and how to repair it. Furthermore, (A. A. Dagade, 2016) proposes a method for managing data duplications, by detecting duplicate records in a single or multiple databases. Following the same concept, (O. Benjelloun, 2008) proposes with Swoosh a two-step technique that matches different tuples to identify duplicates and merge the duplicate tuples into one. The solution proposed for detecting and repairing dirty data in (A. C. Gohel, 2017) resolves errors like inconsistency, accuracy, and redundancy, by treating multiple types of quality rules holistically. In Bleach (Tian & al., 2017) a rule-based data cleaning technique is proposed where a set of rules defines how data should be cleaned.

The cleaning component of PolicyCLOUD adds to the current state-of-the-art an overall data cleaning approach that can be adapted and automatically-adjusted to the severity of the domain and the context of the ingested data. The domain is semantically identified following the semantic meaning and interpretability of the ingested data's content following the bag of words paradigm: a representation of text that describes the occurrence of words within a document. The cleaning process is then adapted accordingly by implementing cleaning actions to answer the relevant needs and requirements. To achieve this, the data cleaning component of PolicyCLOUD detects and corrects (or removes) inaccurate or corrupted datasets, which may contain incomplete, incorrect, inaccurate, or irrelevant data; the component can then replace, modify, or delete those data (also known as "dirty" data). More specifically, the main goals of this component are to (i) ensure a substantial level of trustworthiness of incoming data, (ii) investigate and develop mechanisms to ensure that ingested data is not duplicated/repeated, and (iii) investigate and develop mechanisms that ensure the information can be provided as needed. To achieve all the aforementioned, the data cleaning component supports various data cleaning actions through three discrete steps, which can be provided as independent services that adapt their functionality based on the specificities and severity of the domain on which the cleaning actions must be performed. All this is done according to the pre-specified requirements. The data cleaning component implements the workflow depicted in Fig. 3.

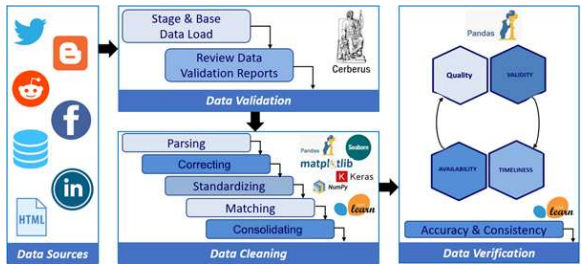

*Fig. 3 Data cleaning workflow*

During the full data cleaning workflow, the data being ingested may be streaming (e.g., Twitter, Facebook, etc) or coming from an originally stored dataset (e.g., webpages, blogs, local datastores, etc). Through the ingestion process, the dataset's domain is identified following the approach of (Kiourtis A., 2019) by discovering and analysing the semantics of the ingested data. As a result, following the process indicated in (A. Mavrogiorgou, 2021), where the domain of the dataset is identified, the required set of data cleaning actions is computed. In the current stage of the PolicyCLOUD project, we are able to differentiate between the various pilot domains, including: radicalization (for RAD), winery (for WINE), smart cities (for SOF), and labour domains (for CAM). Nevertheless, the overall process is already trained to consider the domains of: healthcare, finance, industry, security, and education. The overall handling of each of the various domains follows a similar approach for each of the use cases, as indicated below. Consequently, we did not face any use case specific difficulties and challenges. In this context, to reduce any domain-generated challenges and blocking issues, the data cleaning process is being continuously trained and improved to identify the semantic nature of each domain with larger success rates. This is done by feeding the overall process with additional training material, in order to include supplementary domain identification.

At this point the pipeline continues as follows:
- Data Validation: The basis for this process is a set of rules specified by the data sources registrants. Each validation rule pertains to one of the attributes of the possibly many entities of the dataset. Each rule is translated into one or many constraints that may be mandatory (e.g., specific value length or data type) or optional (e.g., value uniformity, cross-field validity). A final list is built, including the cleaning (corrective) actions to be applied (e.g., deletion, replacement or prediction of a value). Consequently, the Data Validation service is able to validate all the different kinds of incoming data identifying errors associated with lack of conformity with a specified set of rules. To successfully complete this process, it implements all the necessary data validation functionalities and constructs necessary validation rules.
- Data Cleaning: The Data Cleaning service performs the necessary corrections or removals of errors identified by the Data Validation service and, depending on the nature of the error performs an automated data cleaning action based on the predefined rules. Hence, this service ensures a dataset's conformity to mandatory fields and required attributes. To successfully perform this process, the Data Cleaning service exploits multiple open-source libraries (i.e., Pandas, Scikit-learn) to implement all the required cleaning functionalities.
- Data Verification: This process checks the data elements of the dataset for accuracy and inconsistencies. The data verification service ensures that all the corrective actions performed by the data cleaning service have been executed in compliance with the design of the data model. This service ensures that the ingested data will be accurately corrected or completed, and that an ingested dataset will be error-free to the greatest extent possible.

## 5.2 Enhanced Interoperability

Like other domains, policy making deals with very different formats, models, and semantics of data. Data interoperability addresses the ability of modern systems and mechanisms that create, exchange, and consume data to have clear, shared expectations for the contents, context, information, and value of these divergent data (New European Interoperability Framework, 2017).

Data interoperability relies on the system's ability to identify structural, syntactical, and semantic similarities between data and datasets, and to render those data/datasets interoperable and domain-agnostic (DAMA, 2009). Another feature of the enhanced interoperability component is its ability to automatically annotate processed data with appropriate metadata and provide Findable Accessible Interoperable and Reusable (FAIR) data (Hasnain, 2018).

In practice, data is said to be interoperable when it can be easily reused and processed by different applications; this allows different information systems to work together and share data and knowledge. Specifically, semantic interoperability is a key enabler for policy makers, as it enhances their ability to exploit big data and improves their understanding of such data (G. Motta, 2016). On top of this, creating efficient and effective policies in terms of good governance, requires modern policy makers to implement techniques, mechanisms, and applications focused on semantic interoperability to increase their performance and enhance their entire policy-making approach (Lyubomir Blagoev, 2019). This is done by extracting and taking into account parameters and information that may not initially be apparent in data/datasets.

Mapping and creating interoperable data depends on methods that provide semantic and syntactic interoperability across diverse systems, data sources, and datasets. The enhanced interoperability component, designed and implemented within the PolicyCLOUD project, relies on data-driven design by using linked data technologies, such as JSON-LD (JSON for Linking Data, n.d.), and standards-based ontologies and vocabularies. This is coupled with the use of powerful natural language processing (NLP) tasks to improve both semantic and syntactic interoperability of the data and datasets (Zheng, 2017).

PolicyCLOUD introduces a novel approach for achieving enhanced interoperability both for data and datasets. We designed SemAI: a generalized and novel Enhanced Semantic Interoperability hybrid mechanism to ease the extraction of valuable knowledge and information (G. Manias, 2021).

SemAI was designed to achieve high levels of semantic data interoperability to help organisations and businesses turn their data into valuable information, add extra value and knowledge, and achieve enhanced policy making through the combination and correlation of several data, datasets, and policies. To this end, SemAI introduces a multi-layer mechanism that integrates two main subcomponents: the semantic and syntactic analysis and the ontology mapping, both depicted in Fig. 4.
The integration of these two subcomponents provides semantic interoperability across diverse policy-related datasets, even between different policy making domains.

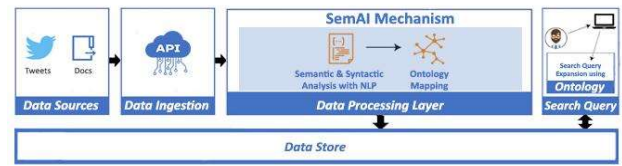

Fig. 4: Enhanced interoperability workflow

We applied SemAI to two families of pilots: RAD and WINE (see Section 1). Since the RAD policy-making and analysis typically uses many datasets, it is critical to interlink and correlate them with annotated interoperable metadata. This permits the discovery of new insights based on merged information that was made interoperable. For instance, when we analysed various datasets, for most of the event records, we were able to extract the radicalization group that was responsible. This enables a deeper understanding of radicalization trends.

For the WINE pilot, we were able to extract and annotate raw tweets by using Named-entity recognition (NER) (Named-entity recognition, 2021) a mechanism for information extraction to locate and classify named entities mentioned in unstructured text into pre-defined categories. The output of NER is named entities along with their role in the tweet, (e.g., Bodegas Viñedos - LOCATION, Campo Cariñena - LOCATION, and San Valero - PRODUCT). We also extracted and annotated topics through the use of several subtasks of the SemAI mechanism, such as topic modelling, topic categorization, part-of-speech (POS), tagging, and NER. By correlating tweets with records from other datasets (e.g., market information about San Valero wine) we were able to derive additional knowledge and insights.

Enhanced interoperability also plays an important role in one of the SOF scenarios, which deals with the analysis of air pollution in the Sofia municipality. This scenario is based on two datasets: the complaints lodged by the citizens with the municipality (the "tickets" dataset) and the IoT records that report pollution measurements as a function of time and location in Sofia. Using SemAI, we aimed to correlate these two datasets to generate a visual analysis of how the actual pollution varies in relation to the opening of air pollution related tickets. This can provide insight, for example, about the level of pollution at which citizens start to open tickets. As a next step, we envision linking this analysis with the (anonymized) medical records of Sofia citizens to discover correlations between improvements in air pollution and the changes in the percentage of Sofia residents who suffer from pulmonary disorders.

## 5.3 Sentiment Analysis

Sentiment analysis is broadly defined as the field of study that examines people's opinions, sentiments, evaluations, attitudes, and emotions based on their written language (Liu, 2012). This field has experienced a tremendous uptake in interest over the last decade in both commercial and research applications due to its applicability in several different domains. Consequently, sentiment analysis is very valuable for learning general opinions about a product, service, or policy. PolicyCLOUD offers sentiment analysis tools to help public administrators and private companies monitor, analyse, and improve their achievements.

Sentiment analysis has matured since its inception in the early 21st century, when it classified long texts into categories according to their overall inclination (B. Pang L. L., 2005). Today, we are seeing remarkable results with the use of neural networks and deep learning techniques, such as convolutional neural networks or recurrent neural networks (L. Zhao, 2021) (C. Sun, 2019) (Manias G., 2020). Statistical techniques, such as discriminative and generative models, (Mesnil, 2014) or supervised machine learning algorithms, such as Naive Bayes, Maximum Entropy Classification and SVMs (B. Pang L. L., 2002) have also been used to classify the different sentiments expressed in written text.

The PolicyCLOUD sentiment analysis component has also evolved, from a document-level approach (Medhat, 2014) (Rachid, 2018) to an entity-level sentiment analysis (ELSA) approach (Colm Sweeney, 2017). The difference between the two approaches lies in the goal of the analysis. In the first version of PolicyCLOUD, our main goal was to understand the general opinion expressed by an author about one main topic (Techniques and Applications for Sentiment Analysis, 2013). In the second version, the goal is to understand the author's opinion regarding various entities at the basic information unit (Karo Moilanen, 2009). This second opinion/sentiment approach can be considered as having an intermediate granularity level between sentence-level sentiment - where the identity of the entity discussed in a sentence is known and there is a single opinion in that sentence - and aspect-level sentiment - where the aim is to extract the sentiment with respect to specific aspects pertaining to the relevant entities (L. Zhao, 2021).

In the document-level approach, we used machine-learning models, such as Vader (Hutto, 2014) to perform sentiment analysis. In the entity-level approach provided by PolicyCLOUD, we used a pre-trained Bidirectional Encoder Representations from Transformers model (BERT model) (Devlin, 2018). An initial pipeline was identified for these activities, which is depicted in Fig. 5 and can be described as follows: (i) the cloud gateway starts the process by providing access to data from a given source (e.g., Twitter); (ii) the data cleaning sub-component performs the initial and necessary pre-processing and cleaning activities on the collected data, (iii) two specific NLP subtasks are executed on the pre-processed/cleaned data: a BERT-based contextual component is triggered for word embedding, and a Named Entity Recognition (NER) component is triggered to extract and classify named entities found in the data, (iv) the enhanced interoperability sub-component annotates the data with information on relevant entities that have already been identified, and on the appropriate topics in which the data have been sorted by topic identification activities, (v) finally, the BERT-based sentiment analysis task is performed, leveraging a ready-to-use Python library (aspect-based-sentiment-analysis 2.0.3, 2021), which has its functionality tied directly to BERT's next-sentence prediction, allowing this task to be formulated as a sequence-pair classification.

The ELSA mechanism enhances the Sentiment Analysis within the PolicyCLOUD project by filtering and providing the corresponding sentiments for identified and extracted entities in the tweets, specifically for the pilot scenario related to Aragon wine marketing policies. The latter can also be used in other use cases such as in the analysis and elaboration of various policies for the Sofia municipality where specific signals from the citizens of Sofia can be processed. In this context, specific sentiments dedicated to a specific topic can be recognized and extracted. For example, a unique signal/post of a citizen can have different sentiment for the transportation issue (e.g., negative) but in parallel it can have the opposite sentiment for the road infrastructure issue (e.g., positive). To this end, the overall sentiment analysis can be enhanced through the ELSA mechanism.

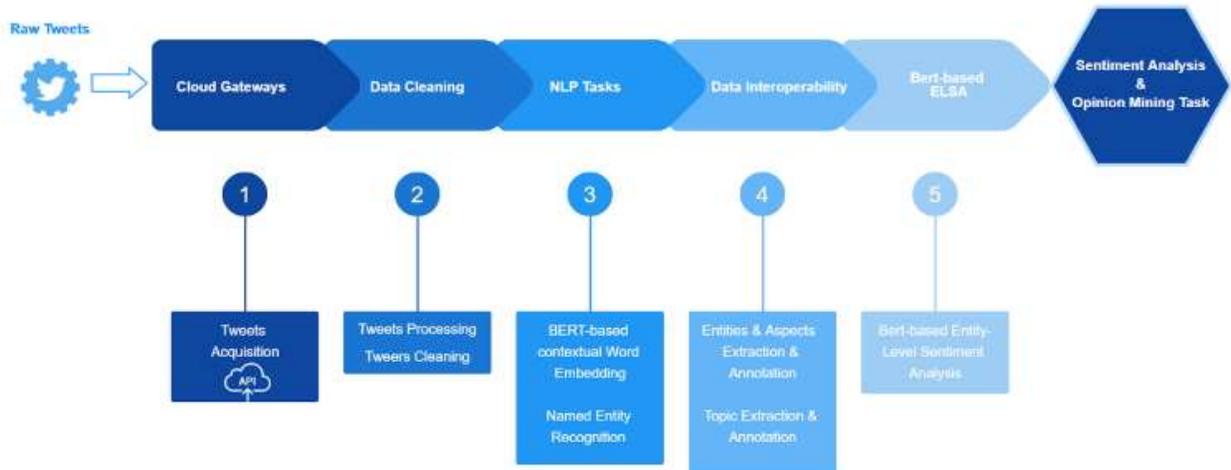
*Fig. 5 ELSA Workflow*

In the following we report our experience applying the sentiment analysis to various pilots of PolicyCLOUD: In RAD, the radicalization related pilot, sentiment analysis is used to extract the sentiment of online activities. This extracted sentiment metadata (e.g., degree of support for violent attacks) permits to annotate the input records (e.g., tweet). In WINE, the wine marketing pilot, SemAI is used to extract from various on-line sources the sentiments related to various wine products. This knowledge can be used in conjunction with the social analytics component to develop and validate marketing policy models and simulations. Moreover, this sentiment analysis annotation can help to understand trends and thus correlate observed changes in sentiment with regards to products acceptance and analyse the impact of various marketing strategies and follow up with prompt reactions when consumer feedback is expressed.

Despite the need for sentiment analysis for the SOF ticket dataset, we were not able to apply our technology since it cannot currently handle non-English texts. This ability has been targeted as one of the future enhancements.

# 6   Social Dynamics Analytics

Social dynamics is the only-non ingest analytics technology developed in PolicyCLOUD, and is used to estimate the social impact of various policies via social simulation. Its goal is to note the possible social implications of various policies with respect to different agent-based models for the populations of interest. Agent-based models and their associated simulation tools have become quite valuable in the analysis of interactions between individuals or groups in social dynamics, as they can capture feedback between the behaviour of heterogeneous agents and their surroundings (Meike Will, 2020). In this paradigm, agents act independently, according to prescribed rules, and adjust their behaviour based on their current state, on that of other agents, and on the environment. Consequently, emergent patterns and dynamics can be observed, arising from local interactions between the agents.

## 6.1   Architecture

The simulation environment includes a user-side and a server-side system. The user-side system allows multiple users to concurrently interact with the social dynamics component through a JavaScript web client interface. Using this interface, they can: specify, edit, or delete a simulation, browse the specifications of simulations stored in the system, execute simulations, upload/download data to a simulation, examine raw simulation results, and compute and visualise simulation analytics. On the server side, all user requests are processed by a web server based on the Phoenix web framework (Phoenix Framework, n.d.). The Phoenix system interacts with three independent components: (1) the simulator built in Elixir (elixir-lang.org), which in turn is built on Erlang (Erlang, n.d.), (2) the analytics component, which includes the meta-simulator environment also built in Elixir, and (3) the storage system. Given that the environment operates as an analytic tool external to PolicyCLOUD, it exposes a REST API through which the PolicyCLOUD environment can receive simulation results in JSON format. The simulator and the rest of the PolicyCLOUD components synchronise their operations by exchanging Kafka messages.

## 6.2 Methodology

The social dynamics component uses agent-based social simulation as its primary analysis tool to evaluate policy alternatives. The policy simulator provides a concurrent environment to manage the state of each individual agent. During each simulation cycle, the simulator spawns a set of concurrent processes - one for each individual agent – and each agent runs its individual and connection dynamics rules and updates its state. Individual rules describe how the attributes of each individual change as a result of the individual's interaction with a set of other individuals. Each such interaction takes place using a connection between the two that has its own attributes. The rules for connection dynamics describe how these connection attributes change over time.

Social dynamics decomposes each policy into a tree hierarchy of goals, objectives, and simulation steps, following the methodology and terminology used in policy analysis. Each goal contains an abstract description of the desired outcomes of a policy. Under each goal hangs a set of alternative objectives that are used to achieve this goal. An objective corresponds to a specific methodology for achieving a goal. Each objective can be decomposed into a sequence of steps, each of which represents a policy execution step in the methodology of the parent objective. We assume that the execution of each step can be simulated, thus providing a value range for its possible outcomes. The social dynamics component simulates each of these steps and embeds a series of analytic tools in the tree hierarchy for a policy; this allows the component to investigate the relationship between simulation outcomes to goals, and operationalize the criteria selected by policy makers. This, in turn, should allow policy-makers to better understand what policy decisions may be recommended in light of their purported goals. Furthermore, by offering a common modelling and execution environment for simulation-based analytics, this component provides a standard basis that facilitates the inspection and comparison of different models for social dynamics.

## 6.3 Social Dynamics Applied to Pilots

In the following section, we describe two pilot cases for the use of social dynamics in policy design. The first pilot is RAD (see Introduction for a short presentation) and provides a qualitative description of the design of a hypothetical and naive policy for containing radicalization. A more detailed description can be found in (N. M. Sgouros, 2021) (Politika, n.d.) (epinoetic, n.d.). We first describe the simulation models we use for modelling policy alternatives, and then show how these models are integrated in a meta-simulation framework that allows their assessment and comparison.

The second pilot, WINE (see Introduction for a short presentation), is concerned with designing a policy to improve consumers' motivation to purchase certain types of wine as compared to their competitors in a specific region. The WINE pilot is being applied to the Aragon region in Spain.

## 6.4 A Simulation Model for Radicalization

### 6.4.1 Background and problem description

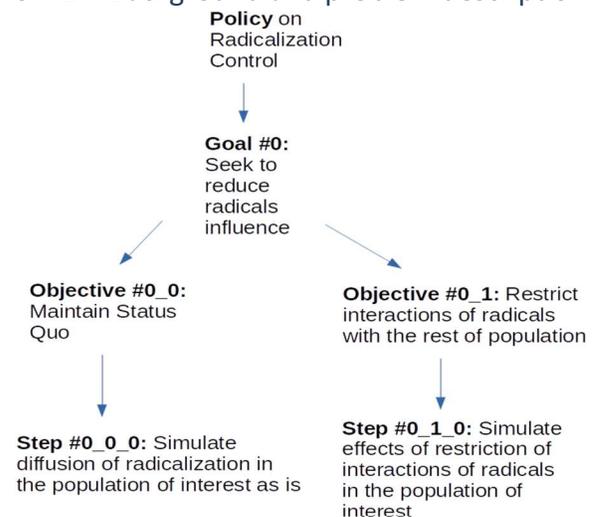

*Fig. 6 Tree hierarchy for Radicalization Policy*

We assume that the radicalization process features the progressive adoption of extreme political, social or religious ideals in the population through social influence. Social influence is defined as change in an individual's thoughts, feelings, attitudes, or behaviours that results from interaction with another individual or a group (Walker, 2015). Social dynamics models a policy's target group as a graph representing a population of autonomous and interconnected agents. We refer to graph nodes as individuals, and to graph edges as their connections. In the RAD model, each individual has an initial `radicalization_status` represented as a real number between -1 (non-radicalized) and +1 (fully radicalised). Each individual can influence other individuals through a number of outgoing, directed connections. Each such connection has:

- A `contact_strength` which indicates whether the individual regards this connection as friendly or not. Contact strength is modeled as a real number between -1 (enemy) and +1 (friend).

- An `influence` representing the level of social influence that a person exerts on other individuals they connect with in terms of radicalization. At each simulation cycle, the `influence` is computed as the product of the `radicalization_status` of the individual and the `contact_strength` of the connection. Therefore, radicalized individuals are expected to influence their friends towards more radicalization, while influencing their enemies towards less radicalization.

During each simulation cycle, individual agents update their current `radicalization_status` by adding to it the sum of the influences they receive through all their incoming edges. At the end of each simulation cycle, the model computes a set of policy-related attributes for the population. Individuals with a `radicalization_status`:

- greater than a defined threshold are considered `radicals`

- less than a conformism-related threshold are considered `conformists` (non-radicals)

- the rest are considered radical `sympathizers`.

Through this example, we compare the social outcomes of applying a policy that restricts the interactions of `radicals` with the rest of the population, with a base case of applying no such policy. Restricting the interaction of `radicals` is modeled as a reduction of the contact strength of their connections with their friends. These friends are those who are the targets of a radical's connections with a `contact_strength` greater than a defined threshold for friendship. Such a reduction is achieved by multiplying the current `contact_strength` with a random coefficient between 0 and 1 during each simulation cycle. Individuals for whom the absolute value of the `contact_strength` of all of their connections is lower than a defined restriction-related threshold are considered to be `isolated` (e.g., they may be under some form of incarceration or surveillance). At the policy level, the `restricted` attribute measures the proportion of isolated individuals in the population.

### 6.4.2 Meta-simulation-based Policy Design for Radicalization

A hypothetical policy to control the spread of radicalization in a population can consist of a tree hierarchy (see Fig. 6), that includes, at its root, one goal (0) to reduce the influence of radicals over the population. Two objectives are analysed with regards to this goal. The first one (0-0) is the base case of maintaining the status quo. Essentially, the objective models the problem that the policy is seeking to address, and provides a base of comparison for the rest of the analysis. The other objective (0-1) seeks to restrict the influence of radicals to the public. Each one of these objectives has a simulation step below it. The 0-0-0 step, under objective 0-0, provides a simulation model of the evolution of radicalization in the population of interest with no policy applied. The other step (0-1-0), under objective 0-1, provides a simulation model of the social dynamics that ensue when a policy seeks to restrict radicals' interaction with the rest of the population.

Each policy has a set of design-specific parameters that can be defined at various levels in the policy hierarchy and follow a top-down propagation in the hierarchy. These include:

1. The population model relevant to the policy, i.e., a graph model that represents the target population in terms of number of nodes and its connectivity patterns (e.g., min/max number of connections per node, use of a random or power-law method for generating the graph);
2. The set of policy-relevant attributes (e.g., the maximum percentage of isolated individuals that can be monitored effectively)
3. The number of rounds and sizes of populations on which each alternative will be tested.

The environment supports a top-down propagation of design parameters in the policy hierarchy. In our example, the values for the population model and policy attributes defined at the policy level are used in the simulation models for both steps 0-0-0 and 0-1-0, while the values for population size and simulation rounds at goal 0 are similarly propagated to both steps 0-0-0 and 0-1-0. The meta-simulation environment automatically generates a bottom-up processing pipeline to transform simulation outcomes of the various alternatives into policy recommendations.

When a user chooses to execute the policy design process hierarchy, then every step in the leaves of this tree runs the simulation model it has been associated with, using the design parameters defined for it. Each step maintains the results of all rounds of simulations it has executed, along with the population size in each one, both indexed under the round number for each. The results of each step are then fed to the objective above it to compute a set of analytics for each of the policy-relevant attributes defined in the design, at this higher level. This set includes the average value of each policy attribute, along with its minimum and maximum values, after all the simulation rounds. Sets of analytics from each objective are then fed to a set of criteria defined by the policy designer at the goal above them. Each criterion evaluates a logical expression involving the policy-related parameters defined for the policy. For example, a criterion can check whether the average value for the `radicals` computed in the objectives below is lower than that of their `sympathizers`, and whether the same average value for their `sympathizers` is lower than that of the `conformists`. After evaluating each objective on the set of criteria defined for the specific goal, the meta-simulator assigns a criteria-ranking map for the objectives based on the proportion of criteria that each one has satisfied. The designer can now consult this map to find out which of the objectives may be preferable for implementing each goal in the policy hierarchy.

## 6.5 Simulation Model for Wine Purchase Motivation

### 6.5.1 Background and problem description

We assume that price and quality are the main factors influencing consumers when purchasing wine. That said, consumers can also be influenced by their exposure to wine-related advertising/marketing campaigns and the wine preferences of their social circle. Based on these assumptions we define the following set of parameters of interest for estimating the purchase motivation for a particular brand of wine (e.g., X) in a specific region:

1. Actual price for X

2. Quality (in a scale of 0 to 1) of X as determined by its average rating in a series of online reviews

3. Estimate of the average price of wines sold in the region of interest

4. Estimate of the maximum price for wine that is acceptable for an average consumer

5. Average quality of the wines sold in the region of interest (0 to 1)

6. Average income of the population in the region of interest

7. Maximum income of the population in the region of interest

8. Average relative exposure of individuals to the advertising campaign for X (0 to 1). We assume that average exposure is proportional to the relative size of the advertising budget for X compared to its competitors

We further assume that the population in the region of interest is represented as a social network, where each node corresponds to an individual. For each individual (e.g., A), each outgoing edge is labelled with a weight representing the influence that A exerts on the wine purchasing decisions of one of its social connections. Each individual has a set of attributes that are relevant towards X. These include A's:

1. Income ranking (in a scale of 0 to 1) as determined by the ratio of its income to the maximum income for the region.
2. Sensitivity to the price of X as determined by the product of the difference of 1 minus A's income ranking times the ratio of the current price of X to the maximum wine price in the region. According to this estimate, poor individuals are more sensitive to the price of wines compared to wealthier ones.
3. Sensitivity to the quality of X as determined by the ratio of the current quality of X to the average quality of wines in the region, times A's income ranking. According to this estimate, wealthier individuals are more sensitive to the quality of wines than poor ones.
4. Susceptibility to the advertising/marketing campaign for X (0 to 1).
5. Susceptibility to social influence towards X (0 to 1)
6. Perceived influence for X from A's social circle. This is computed as the average purchase influence for X stemming from its social circle.

Based on these attributes, the model estimates A's purchase motivation for X as a linear combination of:

1. A's price sensitivity for X.
2. A's quality sensitivity for X.
3. The product of A's advertising susceptibility for X to the intensity of X's advertising campaign.
4. The perceived influence for X for A's social circle.

### 6.5.2 Meta-simulation-based Policy Design to Improve the Purchase Motivation of a Specific Brand of Wine

The purpose of this pilot is to identify changes in the parameters for X (price, quality, and advertisement campaign) that can increase consumers' motivation to purchase it as opposed to its competitors in a specific region. To this end, the designer can simulate various policy alternatives with different values for price, quality, and advertising in order to discover the mix that could improve X's average purchase motivation in the population with respect to its competition. Alternatives can be evaluated and ranked using criteria related to the difference in average purchase motivation for X in the population relative to a base case value of such a motivation. Future policy models will evaluate strategies to increase the purchase motivation for a portfolio of brands relative to their competitors.

## 7 Critical Evaluation

In this section we provide a critical assessment of the PolicyCLOUD framework and its integration with policy pilots.

One of the goals pursued by the PolicyCLOUD framework is to promote transparency in the policy making process.

Transparency has many aspects and a fully transparent process would ideally permit an external observer to be confident that the datasets and analytic functions used in a specific policy process are adequate.

In the following we discuss some of the transparency issues that we encountered in PolicyCLOUD, as well as our solutions.

a) What is the quality of the datasets used? For instance, did they introduce a bias in the analysis done in the process?

b) If multiple datasets are processed jointly in relation to some policy studies, how do we know if they are mutually consistent?

c) Could alternative datasets have been chosen in place of the one(s) used for a policy study? What is the rationale for this choice?

We will take bias, or the lack of it, as an important quality data aspect:
In subsection 4.1, we presented the PolicyCLOUD requirement of providing bias management documentation when registering datasets or functions. The goal is to prevent the uninformed use of biased artefacts.
We noticed that this documentation requirement was not fully satisfactory. The main reason being that a dataset will be detrimental to policy-making only relative to the goals of the intended policy. For instance, let's assume a dataset of recorded nutrition disorders observed in male patients. Obviously, the use of this dataset for elaborating general policies will suffer from an acute gender bias; however, using the very same dataset for policies intended for males will be free of gender bias.
We handled this problem by requiring that dataset documentation specify basic statistical facts out of which one may infer whether it is biased for a given usage (e.g., 87% of the records pertain to males while only 13% pertain to females, in addition 99% of the records relate to individuals living in US towns of more than 100,000 inhabitants).

We identified an additional difficulty that relates to inter-dataset consistency. The problem is the inconsistency of seemingly identical concepts that appear in schemas of different datasets.
For example, should people discouraged by their inability to find suitable employment (e.g., single mothers), and/or those who no longer seek a job, be considered unemployed? Various datasets dealing with "unemployed" will be inconsistent if their answers to this question differ.

Our basic approach for alleviating this problem is to require that the party registering the dataset will document (within the bias documentation) the specificities of the concepts used within the schema of the dataset (e.g., detail whether "unemployed" includes discouraged people).

Our recommendation, however, is to transform the datasets to remove the inconsistency and render them fully comparable. In the employment example, the data augmentation would consist of defining clear concepts and further transform the datasets to address these concepts. This could be done, for example, by assuming separate "unemployed" and "discouraged" statuses, checking if the "unemployed" status of a given person's record should be kept or transformed into a "discouraged" status. This transformation may be inferred from attributes of the record.

Another difficulty is that the choice of a dataset and the specific analytic tool used may not be clearly grounded. In extreme cases, a dataset and their specific processing may be chosen such that they lead to an a-priori chosen policy. In some cases, even if a dataset should clearly be chosen as the basis of the policy making, the analysis may have been applied to a specific subset that will lead to certain conclusions or it can be based on unreasonable assumptions, over-simplification, and problematic estimates of important parameters (Aodha L.., 2017). This can be the case since there is always an ad-hoc character in policy making, which is essentially a political process and therefore reflects political opinions when framing and solving problems (Cairney, 2021). For example, even deciding whether or not a certain level of unemployment is a problem that should be addressed by a policy is arguable as there are views arguing for a 'natural' level of unemployment in an economy. As a first step in alleviating the problem, we require that the policy maker will document answers to a list of questions such as: Why were these specific datasets chosen? Are there other known datasets that could have been chosen? Another proposal is to provide digital tools that promote critique and consensus building in policy design. Allowing people to comment on the platform about the datasets and functions used in different policy cases and even rate them can help to reconcile objective knowledge coming from observational data and/or simulations with its subjective interpretation as is often the case in the social domain (Carlo Martini, 2014). Also providing interfaces that facilitate side-by-side comparisons between different policy alternatives can enhance the level of policy debate.

As a last observation, the effectiveness of a platform such as PolicyCLOUD depends on a redesign of the general policy practice around the use of such open and transparent technologies for public-interest and that may require brave political decisions.

## 8 Related Work

We describe a number of approaches that have been proposed in the context of evidence-based policy making to provide tailored information and guidance for policy practitioners. In this vein, artificial intelligence (AI) can contribute towards more efficient policies. For example, analysis of self-driving datasets (Jiang, 2015) should be the basis of new driving policies that will be needed for this phenomenon. Representative examples include Society 5.0, a project in Japan that aims to analyse data from heterogeneous sources to create evidenced-based policies and provide a human-centric sustainable society (Y. Shiroishi, 2018). Another example is the Foundations for Evidence-Based Policymaking Act introduced in the U.S. to provide better access to high-quality data driving evidence-based policies for federal agencies, government officials, and constituents (Informatica, 2019). Another relevant body that exploits evidence-based policy making techniques is the American Council for Technology and Industry Advisory Council (actiac, n.d.). It facilitates the requirements of evidence-based policies through a mature data management framework, complemented with assessment techniques for managing specific key performance indicators relevant to policies.

The use of big data and AI techniques on massive governmental datasets holds great promise for unleashing innovation and improving public management (J. Hochtl, 2016), to extend government services, solicit new ideas, and improve decision-making (John Carlo Bertot, 2012).
This has triggered attention for NLP and other AI tools that serve as a means for public administration to automatically collect and evaluate citizens' opinions about policies (Boonthida Chiraratanasopha, 2019). Another popular AI technique used in evidenced-based policy making is data mining or "knowledge discovery in databases" which open-endedly looks for patterns in existing data to help policy makers better understand and extract patterns and knowledge (A. Androutsopoulou, 2018).
Several projects propose solutions to deliver end-to-end solutions across the full data path for policy management. The DUET (DUET, n.d.) project proposes the use of digital twins (related to cities systems) through a 3D interface for policy impact exploration and experimentation. A similar approach is proposed by the IntelComp project (Intelcom, n.d.), which tackles the full policy lifecycle, i.e., agenda setting, modeling design, implementation, monitoring and evaluation, through a living-labs approach to involve all relevant stakeholders. The outcomes of the project target different domains with an emphasis on climate change and health.
Looking more towards the actual transition of public authorities towards evidence-based and co-creation policy making, the DECIDO project (DECIDO, n.d.) focuses on the identification of a set of pathways, recommendations and a sound business plan for public authorities.
AI4PublicPolicy (ai4publicpolicy, n.d.) offers an open virtualized policy management environment that provides policy development/management functionalities based on AI technologies, while leveraging citizens' participation and feedback towards re-usability of policies.

While the aforementioned projects also target evidence-based policy making, the PolicyCLOUD project distinguishes itself by (1) focusing on the provision of tools to support policymakers in the collection, aggregation and specialized analysis of heterogeneous datasets, which themselves may be retrieved from heterogeneous sources (the evidence around which policies are to be developed), (2) allowing for visualization of insight extracted from data analysis and simulation of policies developed around such insight, and (3) having designed and developed these functionalities around legal and ethical constraints applicable to datasets and analytical tools, to provide greater assurances of lawfulness and trustworthiness of the PolicyCLOUD platform and its output to users and society at large.

## 9  Conclusion and Future Work

We have presented PolicyCLOUD, a pioneering cloud-based architecture dealing with policy making and its current implementation. We demonstrated how PolicyCLOUD enables organizations both to simply register analytic tools and datasets and to reuse them. We detailed the three ingest analytic tools that were developed to date as well as the modeling and simulation analytical tool.

While the initial results of PolicyCLOUD are very encouraging, our evaluation shows that important capabilities still need to be added. First, we need a way to evaluate the infrastructure financial costs associated with the processing of a new dataset, given its size, its ingest analytical sequence of tools, etc. Another feature is to understand how the reuse of a registered analytic tool can benefit its owner and how to define the liability of these reused tools. In addition, it will be critical to ensure that registered analytical tools will not constitute security problems.

Our initial use of PolicyCLOUD suggests further steps. For example, in the radicalisation pilot mentioned at the end of the Sentiment Analysis subsection 5.3, the location and degree of identification of the tweet author to various social circles of interest can be estimated and thus permit the analysis of radicalization trends as function of location. A next natural step would be to correlate between radicalization trends and radicalization policies in a given country to help policy makers understand the actual impact of the policies and infer any adjustments needed. Also, several technical improvements such as enabling Sentiment Analysis to non-English scenarios will be added to widen the applicability of PolicyCLOUD to new scenarios.

The possibility of enabling user feedback is also important when it comes to encouraging reuse. Hence, we should give the PolicyCLOUD users the opportunity to comment and rate both the framework itself and the reused analytical tools. Last but not least, the PolicyCLOUD platform has been developed subject to various legal and ethical requirements, with the goal of ensuring the platform's lawfulness and maximizing its trustworthiness (and that of the policies generated through it). This includes an innovative dataset/function registration process which, through the requirements imposed upon registrants, should allow platform users to make informed and balanced decisions about the datasets and functions they wish to leverage in their policymaking process. It will be interesting and important to further analyse how diverse PolicyCLOUD users manage these complex requirements through the platform.

## 10  Abbreviations table

| Abbreviation | Signification |
|---|---|
| ABAC | Attribute Based Access Control |
| AI | Artificial Intelligence |
| BERT | Bidirectional Encoder Representations from Transformers |
| DAA | Data Acquisition and Analytics layer |
| HPC | High Performance Computing |
| NER | Named Entity recognition |
| NLP | Natural Language Processing |

**Acknowledgments.** This article focuses on the Cloud Analytics framework of the PolicyCLOUD project which was developed by a subset of the PolicyCLOUD team. The authors are grateful to the rest of the PolicyCLOUD crew for their cooperation and support.

**Funding statement.** "The research leading to the results presented in this article received funding from the European Union's funded Project PolicyCLOUD under grant agreement no 870675. The funder had no role in study design, data collection and analysis, decision to publish, or preparation of the manuscript."

**Competing interests.** `None'

**Data availability statement.** We present some of the core features and functionalities of the PolicyCLOUD platform. To illustrate them, we provide examples that refer to datasets identified by use cases from our partners in the consortium. These partners are themselves policymakers working in specific scenarios; these scenarios were used by the consortium to identify platform requirements and ensure that development of the platform was done around actual and concrete policymaker needs.
The datasets mentioned in this article include
(1) the datasets used in the RAD pilot which include public data derived from the Global Terrorism Database (accessible at: https://www.start.umd.edu/gtd/) and the RAND Database of Worldwide Terrorism Incidents (accessible at: https://www.rand.org/nsrd/projects/terrorism-incidents.html),
(2) the datasets used in the WINE pilot, which derive data from the Twitter social media platform, under the terms and conditions applicable to Twitter's API (available at: https://developer.twitter.com/en/docs/twitter-api),
(3) the datasets used in the SOF pilot, which are compiled by the respective partner as a local public authority (and are thus not publicly available),
(4) the datasets used in the CAM pilot, which include data derived from Open Data Camden (accessible at: https://opendata.camden.gov.uk/).
As these datasets are used merely for illustrative purposes in this article - which is of a translational, rather than research, nature - we do not present in this article any relevant findings stemming from these datasets which would merit verification via data availability.

**Author contributions.** All authors approved the final submitted draft.
Author contributions using the CRediT taxonomy[6] roles as a guide:

Conceptualization: O.B

Data curation: A.M

Funding acquisition: D.K

Methodology: M.T.B, N.S, O.B, P.K, Y.M

Project administration: D.K, O.B, Y.M`

Software: A.K, A.M, G.M, M.S, N.S, O.F, P.K

Supervision: Y.M

Writing – original draft: N.S, O.F

Writing – review & editing: A.K, A.M, D.K, G.M, M.T.B, M.S, N.S, O.F, Y.M

---

[6] https://www.casrai.org/credit.html